\documentclass[sigconf,natbib=true,noacm=true]{acmart}
\usepackage{csquotes}
\usepackage{paralist}
\usepackage{pifont}
\usepackage{makecell}
\usepackage{siunitx}
\usepackage{threeparttable}
\newcommand{\cmark}{\ding{51}}%
\newcommand{\xmark}{\ding{55}}%
\usepackage{listings}
\usepackage{tikz}
\usepackage[htt]{hyphenat}
\usepackage{xcolor,tcolorbox}
\usetikzlibrary{decorations.pathmorphing}
\makeatletter
\newenvironment{btHighlight}[1][]
{\begingroup\tikzset{bt@Highlight@par/.style={#1}}\begin{lrbox}{\@tempboxa}}
{\end{lrbox}\bt@HL@box[bt@Highlight@par]{\@tempboxa}\endgroup}

\usepackage{enumitem}

\newlist{questions}{enumerate}{2}
\setlist[questions,1]{label=RQ\arabic*.,ref=RQ\arabic*}
\setlist[questions,2]{label=\thequestionsi.{\arabic*.},ref=\thequestionsi(\arabic*)}

\definecolor{venngreen}{HTML}{CCE6CC}
\definecolor{vennred}{HTML}{FFCCCC}
\definecolor{vennblue}{HTML}{CCCCFF}

\definecolor{prettyblue}{HTML}{3465a4}
\definecolor{prettyyellow}{HTML}{fce94f}
\newcommand\btHL[1][]{%
  \begin{btHighlight}[#1]\bgroup\aftergroup\bt@HL@endenv%
}
\def\bt@HL@endenv{%
  \end{btHighlight}%
  \egroup
}
\newcommand{\bt@HL@box}[2][]{%
  \tikz[#1]{%
    \pgfpathrectangle{\pgfpoint{1pt}{0pt}}{\pgfpoint{\wd #2}{\ht #2}}%
    \pgfusepath{use as bounding box}%
    \node[anchor=base west, fill=orange!30,outer sep=0pt,inner xsep=1pt, inner ysep=0pt, rounded corners=3pt, minimum height=\ht\strutbox+1pt,#1]{\raisebox{1pt}{\strut}\strut\usebox{#2}};
  }%
}
\makeatother

\newcommand{\tikzcircle}[2][red,fill=red]{\tikz[baseline=-0.5ex]\draw[#1,radius=#2] (0,0) circle ;}%

\lstdefinestyle{patch}{
    basicstyle=\ttfamily\scriptsize,
    moredelim=**[is][{\btHL[fill=green!30]}]{`}{`},
    moredelim=**[is][{\btHL[fill=red!30]}]{@}{@},
    moredelim=**[is][{\btHL[fill=prettyblue!30]}]{??}{??},    
    escapeinside={(*}{*)},
    numbers=left,
    keywordstyle=\ttfamily\bfseries,
    stringstyle=\ttfamily\itshape
}

\definecolor{resultboxgray}{gray}{0.98}

\tcbuselibrary{skins}
\tcbset{shield externalize} 

\newtcolorbox{answerbox}[2][]{
    blanker,
    left=3mm,
    right=3mm,
    borderline west={1.2pt}{0pt}{black},
    title={#2},
    fonttitle=\bfseries,
    coltitle=black,
    #1}

\setcopyright{none}
\acmISBN{}
\acmDOI{}
\acmYear{}
\acmVolume{}
\acmArticle{}
\acmMonth{}
\renewcommand\footnotetextcopyrightpermission[1]{}

\AtBeginDocument{%
  \providecommand\BibTeX{{%
    \normalfont B\kern-0.5em{\scshape i\kern-0.25em b}\kern-0.8em\TeX}}}
\usepackage{hyperref}

\begin{document}

\title{Out of Context: How important is Local Context in Neural Program Repair?}

\author{Julian Aron Prenner}
\email{prenner@inf.unibz.it}
\affiliation{%
  \institution{Free University of Bozen/Bolzano}
  \city{Bozen/Bolzano}
  \country{Italy}
}

\author{Romain Robbes}
\email{romain.robbes@u-bordeaux.fr}
\affiliation{%
  \institution{Univ. Bordeaux, CRNS}
  \city{Bordeaux}
  \country{France}
}

\renewcommand{\shortauthors}{Prenner and Robbes}

\begin{abstract}
Deep learning source code models have been applied very successfully to the problem of automated program repair.
One of the standing issues is the small input window of current models which often cannot fully fit the context code required for a bug fix (e.g., method or class declarations of a project). Instead, input is often restricted to the local context, that is, the lines below and above the bug location.
In this work we study the importance of this local context on repair success: how much local context is needed?; is context before or after the bug location more important? how is local context tied to the bug type?
To answer these questions we train and evaluate Transformer models in many different local context configurations on three datasets and two programming languages. Our results indicate that overall repair success increases with the size of the local context (albeit not for all bug types) and confirm the common  practice that roughly 50-60\% of the input window should be used for context leading the bug.
Our results are not only relevant for researchers working on Transformer-based APR tools but also for benchmark and dataset creators who must decide what and how much context to include in their datasets.
\end{abstract}



\maketitle
\pagestyle{plain}

\section{Introduction}

Deep learning-based methods have shown promising performance in Automated Program Repair (APR), leading to the subfield of Neural Program Repair (NPR)~\cite{chenSequenceRSequencetoSequenceLearning2021, lutellierCoCoNuTCombiningContextaware2020, xiaLessTrainingMore2022a, prennerCanOpenAICodex2022a}. NPR models are trained in such a way that, given a snippet of buggy code as input, the model outputs fixed code. This input is usually comprised of two parts: (I) the buggy code itself and (II) the surrounding \emph{local context} code, that is code that comes before the bug location (pre-context) and after (post-context). The bug location, together with its pre-context and post-context we refer to as the \emph{context window}.

\paragraph{Importance of context} The significance of the local context is twofold. First, it identifies the purpose of a given code snippet, giving the model implicit hints as to how the bug should be fixed. Second, local context is an important source of repair ingredients, that is, code elements relevant for a fix (e.g., variable, field or methods names).
Section~\ref{sec:background} provides more background on context, context variants, and its relation to datasets and NPR models. 

\begin{figure}[htbp]
\centering
\includegraphics[width=\columnwidth]{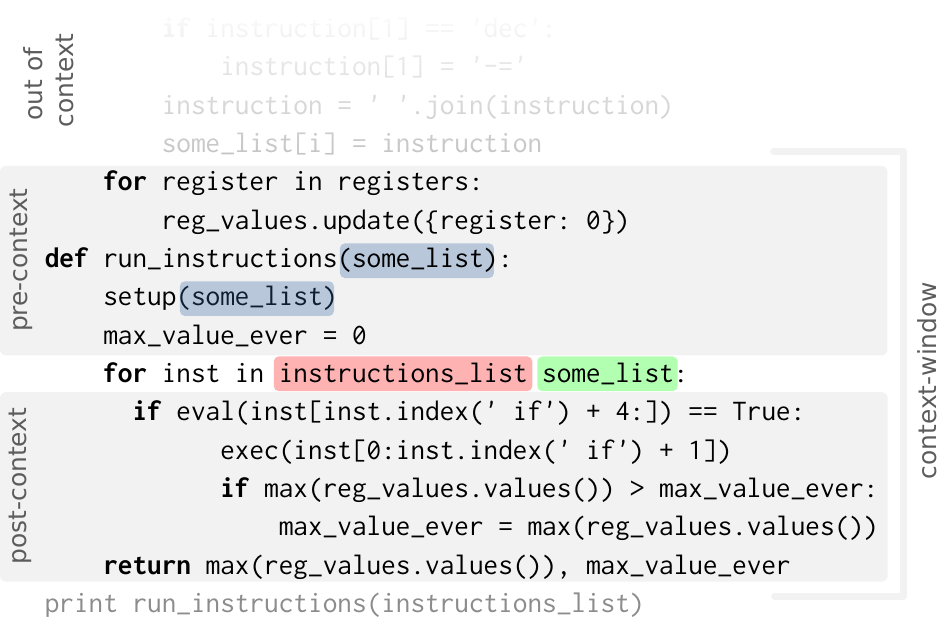}
\caption{A patch from the TSSB-3M~\citep{richterTSSB3MMiningSingle2022} dataset. To fix the bug, the buggy token {\btHL[fill=red!30]\texttt{instructions\_list}} has to be replaced by the correct token  {\btHL[fill=green!30]\texttt{some\_list}}, found in the context ({\btHL[fill=prettyblue!30]\texttt{some\_list}}). The context window has a total size of 10 lines (5 lines of pre-context and 5 lines of post-context); it excludes the first 4 lines and the last line in the figure, which are not part of the model input. Note: we shortened the \texttt{inst} identifier to improve readability.
}
\label{fig:patch}
\end{figure}

\paragraph{Example} Figure~\ref{fig:patch} shows an example. To fix the bug in the presented snippet, an APR model should replace the identifier {\btHL[fill=red!30]\texttt{instructions\_list}} with the identifier {\btHL[fill=green!30]\texttt{some\_list}}. Luckily, this identifier appears within the context ({\btHL[fill=prettyblue!30]\texttt{some\_list}}) and the model is able to pick it up and correctly fix the bug. However, it is easily imaginable that there exist cases in which important identifiers (and other important fix-related knowledge) fall outside the provided context (out of context), especially if the window is small. 

\paragraph{This work}
So far, little work has investigated the importance of such local context in NPR and its effect on repair success. How many lines of context code do we actually need? Should we prefer context before, or after the bug location? Are there bug types that require more or less context?
To answer these and more questions we train and evaluate multiple variants of a Transformer-based NPR model that \emph{can} leverage a large context. Section~\ref{sec:methodology} details our methodology to answer the following research questions:
\begin{questions}
\item \emph{How important is local context for repair success?} We study multiple context sizes, ranging from a single line up to 28 lines on both sides (56 lines) on three datasets (MegaDiff~\citep{monperrusMegadiffDataset600k2021}, TSSB~\citep{richterTSSB3MMiningSingle2022}, and ManySStuBs4J~\citep{karampatsisHowOftenSingleStatement2020}, see Table~\ref{tab:datasets}), totalling several hundreds of thousands of bugs.
\item \emph{How do different bug types and complexity (number of changes) respond to different context sizes and context window positions?} Both, MaySStuBs4J~\citep{karampatsisHowOftenSingleStatement2020} and TSSM-3M~\citep{richterTSSB3MMiningSingle2022} classify bugs into several bug types or bug patterns. We use this bug type labels to analyze how context size affects repair success for bugs of different types. We perform a similar analysis also for the number of changes of a bugfix.
\item \emph{What is the optimal context window position? In other words, given a fixed context budget, how should it be divided among pre-context and post-context?} We experiment with six different context window positions (for four different context window sizes), from only pre-context over several combinations to only post-context.
\item \emph{Is there a connection between the model size (number of parameters), the number of sampled fix candidates and context?}
With more context, the amount of fix ingredients increases. We hypothesize that in order to \emph{fully exploit} context, model size should increase, as should the number of samples. We investigate if this is indeed the case.
\end{questions}

\paragraph{Findings} 
We find that the number of context lines strongly influences repair success, leading to relative improvements up to 29\%. We observe this for context sizes up to over 50 lines (significantly larger than the current practice), however performance varies with bug types and change sizes.
As for the optimal context window position, we observe allocating roughly 50-60\% for pre-context and the rest for post-context yields the best performance; in addition, ensembling multiple contexts improves performance.
Finally, models with more parameters or samples benefit similarly from increased context. Section~\ref{sec:results} details our findings.

\paragraph{Implications} We hope that our work will help NPR researchers to make the most of their models by including a sufficient amount of local context, and clearly document their context choices.
Moreover, we call on dataset creators to include more local context in their datasets. As part of this study, we re-mined the TSSM and ManySStuBs4J datasets, as the the original datasets did not provide enough context (we will release the subset of bugs used in this work with a larger context size as part of our replication package.). We discuss further challenges and opportunities with increased context in Section~\ref{sec:implications}, before discussing the limitations of our work (Section~\ref{sec:limitations}) and concluding (Section~\ref{sec:conclusions}).

\section{Background and Related Work}
\label{sec:background}

Context is an important factor for repair success in NPR~\citep{lutellierCoCoNuTCombiningContextaware2020}. For one, context may act as an abstract \enquote{description} of the code. For example, in Figure~\ref{fig:patch}, identifiers such as \texttt{instruction\_list} and \texttt{register} indicate that this code may be related to a register-based virtual machine implementation. This \enquote{description} may help the model to find and apply the correct fix. Second, context is an important source for ingredient code. Again referring to Figure~\ref{fig:patch}, we see a variable misuse bug. Instead of the correct identifier {\btHL[fill=green!30]\texttt{some\_list}}, the wrong identifier {\btHL[fill=red!30]\texttt{instructions\_list}} is used.
The correct variable name {\btHL[fill=prettyblue!30]\texttt{some\_list}} does appear in the pre-context while the \enquote{buggy} identifier {\btHL[fill=red!30]\texttt{instructions\_list}} appears neither in the pre-context nor the post-context.
It is easy to see that bugs where such ingredients do not appear inside the local context, either because they happen to be far away from the bug location or because the context window is too small to include them, are very hard to fix.

\subsection{Context in APR and NPR}
\label{sec:context}
\label{sec:beyond}
\paragraph{Context as Source for Ingredients}
In a previous study, \citet{yangWhereWereRepair2021} have identified several levels of ingredient code origin, among them most importantly: 
\begin{inparaenum}[I)]
\item intrinsic ingredients, that is ingredients implicitly coming from the specifications of the programming language. In Figure~\ref{fig:patch} these would include Python keywords (e.g., \texttt{def}, \texttt{for} or \texttt{if}) or builtin functions such as \texttt{eval} or \texttt{max}).
\item \emph{method level} ingredients, which includes code elements from the method definition that contains the bug. In Figure~\ref{fig:patch}, {\btHL[fill=prettyblue!30]\texttt{some\_list}} is a method level ingredient.
\item ingredients on the \emph{file or class level}, that is, code located in the same file or class as the bug location (e.g., in Figure~\ref{fig:patch} parts of the preceding function definition appear in the pre-context)
\item donor code coming from the surrounding \emph{package or module},
and finally 
\item ingredients on the \emph{program/project level}, which includes a project's or program's entire codebase.
\end{inparaenum}

\paragraph{Use of local context}
While ingredients on the class and project level are often crucial for a successful repair~\citep{yangWhereWereRepair2021}, they might spread over thousands of lines of code.
Traditional generate-and-validate methods have been able to exploit project level ingredients~\citep{gouesGenProgGenericMethod2012, sahaElixirEffectiveObjectoriented2017, yuanARJAAutomatedRepair2020}. However, so far, NPR systems have been quite limited in the amount of code they can consume. For instance, \citet{chenSequenceRSequencetoSequenceLearning2021} use the surrounding method code as model input, but only experiment with short methods of 50 and 100 tokens of length. Using the method as a context boundary was also done in CoCoNuT~\citep{lutellierCoCoNuTCombiningContextaware2020} and later CURE~\citep{jiangCURECodeAwareNeural2021a} as well as in more recent work on large language models in APR~\citep{xiaAutomatedProgramRepair2023, jiangImpactCodeLanguage2023}. 
In general, we find that previous work often lacks detail about context handling. For instance, RewardRepair~\citep{ye2022neural} is said to use 10 lines of context, but not whether this means 10 lines in total, or for each side. Similarly, since methods can substantially vary in size, the amount of context will be highly variable. 

\paragraph{Context enrichment}
To give the model access to information often not found in the local context, model input may be \enquote{enriched} with further information, in particular with carefully selected file or project level context. For instance, SequenceR~\citep{chenSequenceRSequencetoSequenceLearning2021} adds, in addition to local context, class level information such as class field declarations and method signature stubs. A similar approach was also used for RewardRepair~\citep{yeNeuralProgramRepair2022}. SelfAPR~\citep{ye2022selfapr} adds diagnostic information such as compiler or runtime errors from test executions to model input. FitRepair~\citep{xiaRevisitingPlasticSurgery2023} uses simple text similarity metrics to select possibly relevant identifiers from out of context code and includes them in the model input.

\paragraph{Code Search and Retrieval}
A series of work investigates the use of code search in APR (e.g., ssFix~\citep{xinLeveragingSyntaxrelatedCode2017}, sharpFix~\citep{xinBetterCodeSearch2019}, or LSRepair~\citep{liuLSRepairLiveSearch2018}).
In a very broad sense, this extends the context to entire code corpora and millions of lines of code. However, code from different projects is unlikely to match the code under repair which necessitates adaptation and translation steps~\citep{xinLeveragingSyntaxrelatedCode2017, xinBetterCodeSearch2019}. 
With \textsc{Cedar}~\citep{nashidRetrievalBasedPromptSelection2023} there exists a  APR system that combines code LMs with retrieval.

\paragraph{Architectural}
CoCoNuT~\citep{lutellierCoCoNuTCombiningContextaware2020} and DLFix~\citep{liDLFixContextbasedCode2020} use architectures with special features to better exploit context information.
Finally, a number of works explores the idea of fine-tuning the model on the project under repair in order to encode relevant project level context directly \enquote{into model weights}~\citep{xiaRevisitingPlasticSurgery2023, ye2022selfapr}.

\paragraph{This work} In this study, we focus on local context, that is $n$ lines of code surrounding the bug location, either before or after. This includes intrinsic, as well as method level ingredients and in some cases ingredients at the class level (if e.g., if neighboring function definitions or field declarations fall within the context). A systematic study of enrichment techniques is left as a future work.

\subsection{Context and Transformers}

\paragraph{Transformer window size}
Many state of the art neural program repair (NPR) models are based on the Transformer~\citep{vaswaniAttentionAllYou2017} architecture~\citep{ye2022selfapr, ye2022neural, fu2022vulrepair, berabi2021tfix, xiaLessTrainingMore2022a}.
One important limitation of this architecture is its limited input window size, which is fixed at training time. Although there have been successful efforts to expand the input window of general Transformer models, recent NPR models use input windows of only several hundreds of tokens. For instance, VulRepair~\citep{fu2022vulrepair}, FitRepair~\citep{xiaRevisitingPlasticSurgery2023} and AlphaRepair~\citep{xiaLessTrainingMore2022a}, RewardRepair are all limited to only 512, SelfAPR~\citep{ye2022selfapr} to 768 model tokens. Often, the amount of context is simply determined by the space left in the input window after the buggy code was placed in it~\citep{xiaLessTrainingMore2022a}.

\paragraph{Position embeddings}
By itself a Transformer has no notion of sequences; it sees its input as a set of elements. The original Transformer architecture added \emph{position embeddings} to give greater weight to a token's neighbour \citep{vaswaniAttentionAllYou2017}. This bias allows it to model its input as sequences. The original position embeddings are \emph{absolute} and tied to the model's input window, causing very poor generalization to longer sequences. Alternative ways to encode token position have been proposed, including relative position embedding~\cite{shawRelativePosition2018}, rotary position embeddings (RoPE~\citep{suRoFormerEnhancedTransformer2022}), and biases to the attention \cite{pressTrainShortTest2022}. These allow fine-tuning to longer sequences, and (to a limited degree) extrapolation to longer sequences without fine-tuning \cite{pressTrainShortTest2022}.

\paragraph{The T5/CodeT5 model in NPR}
T5~\citep{raffelExploringLimitsTransfer2020} and
CodeT5~\citep{wangCodeT5IdentifierawareUnified2021} a variant specifically pre-trained for code-related tasks are popular Transformer models for sequence-to-sequence (seq2seq) tasks. Since program repair can be naturally represented as a seq2seq task by letting the input sequence be the buggy code and the output sequence the fixed code, this model has been used extensively in the recent NPR literature. \citet{berabi2021tfix} fine-tune T5 on coding errors provided by ESLint. \citet{ye2022selfapr, ye2022neural} use CodeT5 as the basis for both, RewardRepair~\citep{ye2022neural} and SelfAPR~\citep{ye2022selfapr}.
\citet{fu2022vulrepair} develop VulRepair, a CodeT5-based Automated
Vulnerability Repair system.
\citet{kimSystematicAnalysisDefectSpecific2022} study the effect of code abstraction techniques in NPR and use T5/mT5 models in all of their experiments.
Similarly, \citet{xiaRevisitingPlasticSurgery2023} use CodeT5 in their analysis of the plastic surgery hypothesis in NPR.

\paragraph{This work}
In addition to its widespread use in NPR, we use CodeT5~\citep{wangCodeT5IdentifierawareUnified2021}  for a very specific reason: its T5 architecture~\citep{raffelExploringLimitsTransfer2020}, unlike most, uses \emph{relative position embeddings}~\citep{shawRelativePosition2018}. This allow us to fine-tune a T5 model with an arbitrary input window size~\citep{pressTrainShortTest2022, raffelExploringLimitsTransfer2020}. To our knowledge, we are the first to fine-tune a model for a larger context window in NPR. For all experiments in this work use an input window size of 1024, that can fit \emph{twice as much} context as most of the models mentioned above.

\subsection{Context and Datasets} \label{sec:datasets}

Most existing datasets used in NPR are large collections of bug fixes mined from code repository commits. In the following, we analyze how context is handled on the \enquote{dataset level}.

\paragraph{BFP}
BFP~\citep{tufanoEmpiricalInvestigationLearning2018} is a collection of over 65,000 bug fixes mined from GitHub. Each dataset sample includes the full method surrounding the bug location, although the dataset focuses on short methods.

\paragraph{CoCoNuT}
In the same fashion, the dataset used by CoCoNuT~\citep{lutellierCoCoNuTCombiningContextaware2020} was mined from open-source project commits and includes context only up to the method boundary.
In particular, the CoCoNuT dataset splits hunks of changes occurring in a single commit into separate dataset instances. Further, whitespace, including newlines is stripped from the dataset instances which makes estimating the context size in terms of number of lines very difficult.

\paragraph{ManySStuBs4J}
ManySStuBs4J~\citep{karampatsisHowOftenSingleStatement2020} is a dataset of over 150,000 single statement Java bugs categorized into 16 bug patterns. We find that  the diffs in ManySStuBs4J have a total length of only 13 lines of code (including the bug itself). The dataset provides the commit hashes and repository identifiers for all dataset instances.

\paragraph{TSSM-3M}
TSSM-3M~\citep{richterTSSB3MMiningSingle2022} is a dataset of over three million single statement Python bugs classified into 20 bug patterns loosely following the categorization of ManySStuBs4J. Our analysis shows that the examples (diffs/patches) in TSSM-3M have a median length of only 9 lines (mean 8.8), that is, less than 4 lines of pre- and post-context. Here too, commit hashes and repository names are provided for all instances.

\paragraph{MegaDiff}
MegaDiff~\citep{monperrusMegadiffDataset600k2021} contains over 660,000 Java diffs with changes ranging from 1 to 40 lines. They come with full file-level context, containing all files affected by changes in a single diff.

\paragraph{This work}
We use the MegaDiff, ManySStuBs4J and TSSM-3M datasets (Table~\ref{tab:datasets}). MegaDiff was chosen because it provides full file level context, ManySStuBs4J and TSSM-3M because they contain all the required information for re-mining. As neither ManySStuBs4J nor TSSM-3M provide sufficient amount of context, we had to re-mine selected subsets with full file-level context (see Section~\ref{sec:remining}). BFP and the CoCoNuT dataset were not considered in this work as they
do not provide full file level context (method level only) nor were we able to find commit information necessary for re-mining .

\section{Methodology}
\label{sec:methodology}

\begin{table}[htbp]
  \footnotesize
  \caption{Datasets used in this work, along with the number of samples used (subset) for training and evaluation.}
  \label{tab:datasets}
  \begin{tabular}{cccccc}
    \toprule
    Dataset & Train & Test & Lang. & Labels \\
    \midrule
    MegaDiff~\citep{monperrusMegadiffDataset600k2021} & 201,358 & 22,479 & Java & \xmark{} \\
    ManySStuBs4J~\citep{karampatsisHowOftenSingleStatement2020} & - & 12,714 & Java & \cmark{} \\
    TSSM-3M~\citep{richterTSSB3MMiningSingle2022} & 424,873 & 46,791 & Python & \cmark{} \\
  \bottomrule
\end{tabular}
\end{table}

To study the effect of local context on repair success, we
\begin{inparaenum}[(I)]
    \item re-mine with full context a subset of examples from existing datasets, which we pre-process for different context size (Section~\ref{sec:remining})
    \item fine-tune various configurations of our model on the corresponding training sets (Section~\ref{sec:training}) and finally
    \item generate and evaluate bugfixes for all bugs in the test set to carry out our experiments (Section~\ref{sec:experiments}).
\end{inparaenum}

\subsection{Re-Mining and Pre-Processing}
\label{sec:remining}

\paragraph{Datasets}
We study three datasets that cover two programming languages: MegaDiff~\citep{monperrusMegadiffDataset600k2021}, ManySStuBs4J~\citep{karampatsisHowOftenSingleStatement2020}, and TSSM-3B~\citep{richterTSSB3MMiningSingle2022}. In addition, ManySStuBs4J and TSSM-3B provide SStuB patterns, that is, bug type labels; we use these to analyze the effect of context size on different bug types in RQ2. For better comparability with the other datasets, MegaDiff\textsubscript{SB} (simple bugs) denotes the subset of MegaDiff bugs that require at most two changes to fix.

\paragraph{Filtering}
Following previous work~\citep{lutellierCoCoNuTCombiningContextaware2020, xiaLessTrainingMore2022a, chenSequenceRSequencetoSequenceLearning2021, liDLFixContextbasedCode2020, ye2022neural}, our study focuses on single-hunk bugs. This is all the more important as the presence of multiple hunks would involve multiple contexts, which would increase the complexity of our study. As such, multi-hunk bugs in MegaDiff were filtered out. Note that we keep examples with multiple changes in a single hunk. We observed that in ManySStuBs4J, some commits included multiple bug patterns and were split in several dataset instances; such cases were also considered multi-hunk edits and consequently filtered out. No filtering was necessary for TSSB-3M which only contains single-statement and thus single-hunk bugs; we focus on a subset of about 450,000 bugs due to limited computational resources. Table~\ref{tab:datasets} shows the size of the datasets after filtering. Due to its final size, we use ManySStubs4J as a test set (with MegaDiff as training set).

\paragraph{Re-mining commits}
As mentioned previously, in our experiments we require a large local context (up to 56 lines). Unfortunately, this is neither provided by ManySStuBs4J nor by TSSM-3B. As repository identifiers and commit hashes are provided in both cases, we re-mine dataset instances with full file level context. This was done using the GitHub API. On rare occasions, commits could not be fetched, possibly because the corresponding repository has been deleted or is no longer public. In contrast to the other two datasets, MegaDiff~\citep{monperrusMegadiffDataset600k2021} is released with full file-level context. Thus, re-mining was not necessary.

\paragraph{Pre-Processing} 

\begin{figure}[htbp]
\centering
\begin{lstlisting}[language=Python,style=patch,numbers=none,xleftmargin=0pt,xrightmargin=0pt,framesep=0pt,morekeywords={CHANGES,CHANGEE}]
def run_instructions(some_list):
    setup(some_list)
    max_value_ever = 0
<CHANGES>
    for inst in instructions_list:
<CHANGEE>
        if eval(inst[inst.index(' if') + 4:]) == True:
            exec(inst[0:inst.index(' if') + 1])
            if max(reg_values.values()) > max_value_ever:
(*\tikz\draw[decorate,decoration={zigzag,amplitude=1.5pt},line width=0.01pt,color=gray!40] (0,0) -- (\linewidth,0pt); *)
<CHANGES>
    for inst in some_list:
<CHANGEE>
\end{lstlisting}
\caption{Unified bug format. Model input above the zigzag line (with three lines of pre and post context), output below. \texttt{<CHANGES>} and \texttt{<CHANGEE>} indicate the buggy code section.  }
\label{fig:unified}
\end{figure}

We strip all empty lines from the dataset examples. To unify the examples from the three datasets we parse the diffs and select context code lines, buggy code lines and the ground-truth (i.e. fixed) code lines; we transform them into the format specified in Figure~\ref{fig:unified}. 
In line with previous work~\citep{ye2022selfapr, lutellierCoCoNuTCombiningContextaware2020, xiaLessTrainingMore2022a}, we assume perfect fault localization, that is, we assume the bug location is known. This avoids confounding of localization and repair performances~\citep{liuYouCannotFix2019}.
We indicate the start and end of the buggy code section with marker tokens (\texttt{<CHANGES>} and \texttt{<CHANGEE>}), similar to SequenceR~\citep{chenSequenceRSequencetoSequenceLearning2021}. In case of multiple changes, each change is marked with tokens and surrounded by its corresponding $n$ context lines. Overlapping context, that is, context lines shared by multiple changes, are fused into single blocks.
The target model output consists of the fixed version of the buggy input lines (without context).

For each configuration in Table~\ref{tab:configs} we generate training test sets where we select the corresponding number of pre/post-context lines. Due to limited computational resources, we skip some TSSM context sizes. In early experiments we noticed diminishing returns above 50 lines of context, which is why we stop at 56 context lines.

\subsection{Training and Evaluation}\label{sec:training}

\begin{table}

  \caption{Context configurations in this study (pre-context lines/post-context lines).}
  \footnotesize
  \label{tab:configs}
  \begin{threeparttable}
  \begin{tabular}{cc}
    \toprule
    \textbf{Datasets} & \textbf{Configurations} \\
    \midrule
    \makecell{MegaDiff~\citep{monperrusMegadiffDataset600k2021}\\ManySStuBs4J~\citep{karampatsisHowOftenSingleStatement2020}} &  1/1-28/28 \\
    TSSM-3M~\citep{richterTSSB3MMiningSingle2022} & \makecell{1/1-16/16, 18/18, 20/20,\\ 22/22, 24/24, 26/26, 28/28} \\
    \hline
    All\tnote{1} & \makecell{5/0, 4/1, 3/2, 2/3, 1/4, 0/5,\\ 10/0, 8/2, 6/4, 4/6, 2/8, 0/10,\\ 20/0, 16/4, 12/8, 8/12, 4/16, 0/20,\\ 30/0, 24/6, 18/12, 12/18, 6/24, 0/30,\\ 40/0, 32/8, 24/16, 16/24, 8/32, 0/40}  \\
    
  \bottomrule
\end{tabular}
\begin{tablenotes}
\item[1] configurations for different context window positions
         (sizes 5, 10, 20, 30 and 40)
\end{tablenotes}
\end{threeparttable}
\end{table}

We implement all of our experiments on top of Huggingface's \texttt{transformers} library~\citep{wolfTransformersStateoftheArtNatural2020}. We fine-tune 
the 60M pre-trained CodeT5 model on the configurations, that is, different datasets and pre-context and post-context sizes, as listed in Table~\ref{tab:configs}. We then generate five fix candidates for each bug in the corresponding test set. The generated model output is evaluated using an exact-match metric.

\paragraph{Hyper-Parameters}
To obtain a fair basis for comparison, we use the same hyper-parameters for all context configurations (unless explicitly specified). We train with FP16 precision, a learning rate of \SI{1e-4}, and a batch size of 12, accumulated over two steps on a consumer-grade NVIDIA RTX 3090 with 24GB of memory. For generation we use beam search with five beams and a maximum length of 1024 generated tokens. 

\paragraph{Evaluation}
We calculate repair success (accuracy) as fraction of examples in the test set where at least one of the five generated fix candidates (i.e. top-5) lexically matches the ground-truth fix, ignoring any whitespace. Exact matching as a metric for repair success was used in previous work~\citep{berabi2021tfix, fu2022vulrepair}.

\paragraph{Tokenization}

Since our model's input window is limited to a 1024 tokens, there is a limit at which further increasing context has no effect, as the exceeding input would be simply truncated away.
With a median token count of 592 and an upper quartile of 699 at 56 lines of context we are well within this limit. At 56 lines of context, truncation is only necessary for 2.2\% of examples. Truncation is done by dropping the last $n$ excessive tokens.

\subsection{Experiments}\label{sec:experiments}

To answer our research questions, we carry out four main experiments:
\begin{inparaenum}[1)]
    \item we train and evaluate repair success for different context sizes (using symmetric context windows),
    \item we further analyze the relationship between repair success, change type, and change size, 
    \item we train and evaluate repair success for different context window positions (i.e., asymmetric context window or different amounts of per-context and post-context) and finally
    \item we train and evaluate repair success for a larger model (220M parameters) and for a larger number of generated fix candidates (small model).
\end{inparaenum}

\subsubsection{Context Size (RQ1)}
For each context size of 1 to 28 lines of code (on both sides) and each of the three datasets (MegaDiff, TSSM-3B and ManySStuBs4J; see Table~\ref{tab:configs} for a list of all context configurations) we fine-tune and evaluate a CodeT5 Transformer as described in \ref{sec:training}.

\subsubsection{Change Type and Change Size (RQ2)}
We analyze the performance of the models from RQ1 to get finer-grained insights. We use the labels of TSSM and ManySStuBs4J to compare the performance of models trained on different context sizes for several types of changes. In the same spirit, we compare how the performance of these models varies with change size on MegaDiff.

\subsubsection{Context Window Position (RQ3)}
\begin{figure}[tbhp]
\centering
\includegraphics[width=\columnwidth]{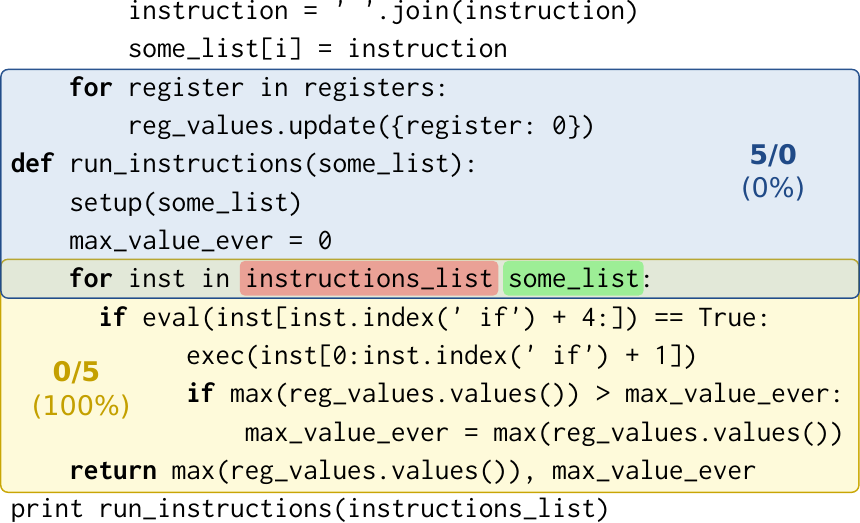}
\caption{The two most extreme window positions 5/0 (0\% -- only pre-context \tikzcircle[fill=prettyblue!30]{3pt}) and 0/5 (100\% -- only post-context \tikzcircle[fill=prettyyellow!30]{3pt}) for a context size of 5. \iffalse In our context window experiments we slide windows of different sizes (5, 10, 20, 30, 40) between these two positions (Table~\ref{tab:configs}).\fi}
\label{fig:window-pos}
\end{figure}

We carry out a series of experiments with asymmetric window sizes.
Here, we keep the context window at a constant size, however, by adjusting pre-context and post-context, the window is slid over the bug location. For example, for a context size of 5 we evaluate at 5/0, 4/1, 3/2, 2/3, 4/1 and 0/5, where $x$/$y$ denotes $x$ lines of pre-context and $y$ lines of post-context. Figure~\ref{fig:window-pos} illustrates this for the two extreme window positions 5/0 and 0/5. All intermediate positions are obtained by \enquote{sliding} the window from top to bottom. The window position can also be described as the percentage of post-context: at position $n\%$, $n\%$ of the window are filled with post-context and $100\% - n\%$ with pre-context (e.g., 0\% for position 5/0, and 100\% for 5/0).
We do this for windows of 5, 10, 20, 30 and 40 lines and 7 sliding positions (0, 20, 40, 50, 60, 80, and 100\%), adjusting the sliding step size accordingly (see Table~\ref{tab:configs} for a list of context configurations). Training and evaluation are performed as outlined above.

\subsubsection{Model Size and Fix Candidate Count (RQ4)}
We hypothesize that a larger model, or a larger number of generated fix candidates, can make better use of larger contexts. We study whether the margins between smaller and larger models (or between different numbers of fix candidates) increase over-proportionally with larger context sizes.

\paragraph{Larger model}
For a selected number of context sizes (1, 7, 14, 21 and 28 lines on both sides) we re-run the context size experiment on a larger version of the CodeT5 model (220M). This experiment is also only carried out for MegaDiff (training and evaluation) and ManySStuBs4J (evaluation only). We double the number of accumulation steps in order to be able to train the larger model with the same (accumulated) batch size of 12 on the same hardware.

\paragraph{More fix candidates}
For all other experiments we use five fix candidates per bug, sampled using beam search. Here, we try 10 and 15 fix candidates per bug (evaluating using top-10 and top-15).
To account for the increased memory requirements when using more beams, we reduce the maximum length of generation from 1024 to 256 tokens. Note that the model only needs to generated the buggy code section, which rarely exceeds 256 tokens. We also enable \emph{early stopping} to speedup the generation process. With early stopping, the search is stopped as soon as $n$ complete solutions have been generated (otherwise the search may continue to find better solutions). For a fair comparison, we also regenerate the top-5 fix candidates with early stopping and a 256 token limit.

\section{Results}
\label{sec:results}
At a glance, we find that:
\begin{inparaenum}[(I)]
    \item overall, more context increases repair success, but not always consistently (Section~\ref{sec:rq1})
    \item the importance of context size strongly depends on the specific bug type and change size (Section~\ref{sec:rq2})
    \item context windows which are centered around the bug location yield the best performance, but extreme windows are useful in ensembles (Section~\ref{sec:rq3}) and
    \item we see no convincing evidence that either larger models or a higher number of fix candidates can better exploit larger context sizes (Section~\ref{sec:rq4}).
\end{inparaenum}

\subsection{Context Size (RQ1)}
\label{sec:rq1}

\begin{figure}
    \centering
    \includegraphics{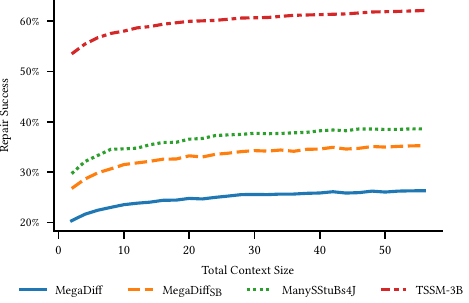}
    \caption{Repair success as a function of context size.}
    \label{fig:rq1}
\end{figure}

We analyze the symmetric context configurations (i.e., $n$ lines of context in the pre-context and the post-context) to estimate the importance of local context on repair success. As shown by Figure~\ref{fig:rq1}, repair success steadily increases as more local context is available.

\paragraph{MegaDiff}
For MegaDiff, repair success increases from 20.4\% with a single line of context to 26.3\% with 28 lines of context on both sides. For simple bugs (MegaDiff\textsubscript{SB}) repair success is considerably higher with numbers ranging from 26.8\% (one line of context) to 35.3\% (28 lines of context).

\paragraph{ManySStuBs4J}
For the ManySStuBs4J dataset, repair success ranges from 29.7\% with single-line context, to 38.6\% with 27 lines of context. Here, repair success does not peak at the maximum context size of 28/28, where performance is slightly lower (-0.008\%).

\paragraph{TSSB-3M}
Overall performance was best for TSSB-3M, where repair success is beyond 50\% for all context sizes.  
Repair success starts at 53\% with a single-line context and reaches 62\% with 28 lines of context on both sides.

\subsubsection{Consistency of Results}
Our results indicate that, \emph{in general}, growing context increases repair success. The ideal case is pictured in Figure~\ref{fig:more-context}: as the correct ingredients enter the context, the model succesfully use them. On the other hand, it was conjectured in previous work that too much context may \enquote{drown} the actual bug in noise, confuse the model and thus stymie repair success~\citep{lutellierCoCoNuTCombiningContextaware2020}. We do observe this, as pictured in Figure~\ref{fig:less-context}. Moreover, upon inspecting our results, we find that the model often \enquote{spontaneously} fails to fix a bug at a certain context size $c$, despite it correctly fixing the bug at one or more lower context sizes. To get more insight on this, we group bugs correctly fixed in at least \emph{some} context size into four categories:
\begin{enumerate}
    \item bugs that are fixed for all context sizes;
    \item bugs that \emph{improve} with context (they are not correctly fixed below a context size $c$, but for all context sizes equal or larger than $c$);
    \item bugs that \emph{degrade} with context (they are \emph{only} be fixed below a context size $c$ and not for context sizes equal or larger than $c$);
    \item \emph{erratic} bugs with more complex patterns (e.g., the model might correctly fix a bug at context sizes 18, 21, or 24, without clear reason of why it failed at, say, context size 20).
\end{enumerate}

We find that 37\% of bugs were fixed for all context sizes (case 1). Bugs with consistent repair success starting from some context size $c$ (case 2, Figure~\ref{fig:more-context}) are surprisingly low with roughly 9\%. On the other hand, we find that the number of bugs that can only be fixed at small context sizes (case 3, Figure~\ref{fig:less-context}) are rare (\textasciitilde 1\%). Finally, the remaining 35\% show some degree of erratic behavior (case 4), which we found to be surprisingly high. In particular, roughly 5\% of the bugs show an \enquote{island pattern}, where a bug can be correctly fixed for a contiguous range of context sizes that is surrounded on both sides by context sizes for which models could not find a fix. The inverse case, that is, a fix can be found \emph{except} for a contiguous block of context sizes in the middle appears with a frequency of about 4.5\%. These inconsistencies indicate that the models lacks robustness; further implication of this are discussed in Section~\ref{sec:implications}.

\newcommand{\xmarkrect}{\tikz[baseline=(X.base)]\node[fill=red!30,rectangle,inner sep=2pt, outer sep=2pt, rounded corners=2pt] (X) {\xmark};}
\newcommand{\cmarkrect}{\tikz[baseline=(X.base)]\node[fill=green!30,rectangle,inner sep=2pt, outer sep=2pt, rounded corners=2pt] (X) {\cmark};}

\newcommand{\xmarkrectt}[1]{\tikz[baseline=(X.base)]\node[fill=red!30,rectangle,inner sep=2pt, outer sep=2pt, rounded corners=2pt] (X) {\xmark\textsubscript{#1}};}
\newcommand{\cmarkrectt}[1]{\tikz[baseline=(X.base)]\node[fill=green!30,rectangle,inner sep=2pt, outer sep=2pt, rounded corners=2pt] (X) {\cmark\textsubscript{#1}};}

\begin{figure}[tbhp]
\centering
\begin{lstlisting}[language=Java,style=patch,numbers=none,xleftmargin=0pt,xrightmargin=0pt,framesep=0pt,morekeywords={CHANGES,CHANGEE}]
(*\cmarkrectt{5}*) game.player.updateMotion(game.player.getPosition(), v, (*\textellipsis{}*));
(*\cmarkrectt{4}*) //other Ben's doing...
(*\cmarkrectt{3}*) if(!v.equals(Vec3.zero))
(*\cmarkrectt{2}*) {
(*\xmarkrectt{1}*)   game.transmitPlayerPosition();
      (**)`transmittedStop = false;`
(*\xmarkrectt{1}*) }
(*\cmarkrectt{2}*) else if(!??transmittedStop??)
(*\cmarkrectt{3}*) {
(*\cmarkrectt{4}*)  game.transmitPlayerPosition();
(*\cmarkrectt{5}*)  transmittedStop = true;
\end{lstlisting}
\caption{Bug from MegaDiff, where an assignment to \texttt{transmittedStop} needs to be added. 
To succeed, the model requires at least two lines of context on both sides (\protect\cmarkrectt{2}). This is likely due to \texttt{transmittedStop} coming into context (\tikzcircle[fill=prettyblue!30]{3pt}). }
\label{fig:more-context}
\end{figure}

\begin{figure}[tbhp]
\centering
\begin{lstlisting}[language=Java,style=patch,numbers=none,xleftmargin=0pt,xrightmargin=0pt,framesep=0pt,morekeywords={CHANGES,CHANGEE}]
(*\xmarkrectt{5}*) public void resume() {
(*\xmarkrectt{4}*) }
(*\xmarkrectt{3}*) public void onDestroy() {
(*\xmarkrectt{2}*)   System.out.println(??"PApplet.onDestroy() called"??);
(*\cmarkrectt{1}*)   super.onDestroy();
      (**)`finish();`
(*\cmarkrectt{1}*) }
(*\xmarkrectt{2}*) //////////////////////////////////////////////////////////////
(*\xmarkrectt{3}*) // ANDROID SURFACE VIEW
(*\xmarkrectt{4}*) SurfaceView surfaceView;
(*\xmarkrectt{5}*) SurfaceHolder surfaceHolder;
\end{lstlisting}
\caption{Bug from MegaDiff where a missing call to \texttt{finish()} needs to be added. This bug is correctly fixed only with a context of one line on both sides (\protect\cmarkrectt{1}). The model is likely led astray by code highlighted in blue (\tikzcircle[fill=prettyblue!30]{3pt})  \protect\footnotemark.}
\label{fig:less-context}
\end{figure}

\footnotetext{The likely culprit here is the string {\btHL[fill=prettyblue!30]\texttt{"PApplet.onDestroy() called"}}: if it is included in the context, the model predicts fixes that include \texttt{PApplet}, such as inserting \texttt{PApplet.onDestroy();}, \texttt{pApplet.onDestroy();} or \texttt{PApplet.onResume();}}

\begin{answerbox}{Answer to RQ1}
Overall, local context has a strong effect on repair success. For MegaDiff \emph{symmetrically} increasing total local context from two lines to 56 lines increases repair success by almost 6\%. For simple bugs (ManySStuBs4J, TSSM-3M and MegaDiff\textsubscript{SB}) performance increased by over 8\%. Given this, we see that local context is an important factor for repair success. However, a large amount of bugs are inconsistently repaired across context sizes.
\end{answerbox}

\subsection{Effect of Bug Type and Change Count (RQ2)}
\label{sec:rq2}
As mentioned previously, the labels (SStuB patterns) in ManySStuBs4J and TSSB-3M allow us to analyze the effect of context size for different label types. Similarly, a number of bugs in MegaDiff have multi-change fixes, making a similar analysis possible for a bug's change count (i.e., the number of changed lines in the diff fixing it).

\begin{figure}[tbhp]
    \centering
    \includegraphics[width=\columnwidth]{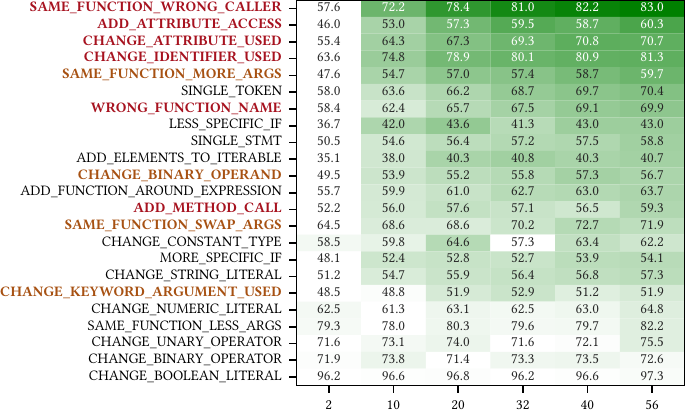}
    \includegraphics[width=\columnwidth]{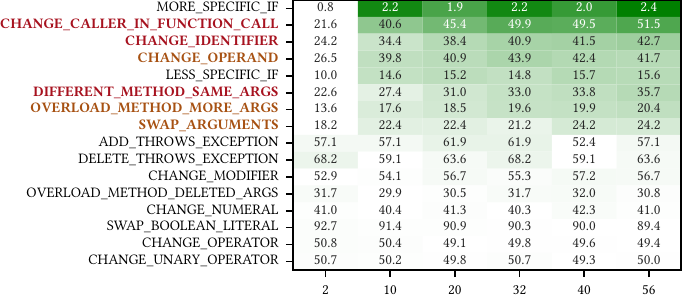}
    \caption{Heatmaps vizualizing the effect of context (x-axis) for bugs of different bug types (SStuB patterns). Top TSSM-3B, bottom ManySStuBs4J. Bug types in red involve identifiers, those in orange likely involve identifiers. Coloring expresses the performance ratio relative to the context size with minimum performance for the corresponding bug type; a ratio of 1.0 corresponds to white.}
    \label{fig:heatmap-labels}
\end{figure}

\subsubsection{Bug Type}

We visualize the effect of context on bug type as a heatmap highlighting relative improvements (Figure~\ref{fig:heatmap-labels}). While some bug types benefit from a larger context, for others, more context hardly makes a change or even slightly lowers performance.

\paragraph{Responding patterns}
For both datasets, we see a very strong effect for the \enquote{change caller} bug types (> 25\% absolute difference in repair success between 2 and 58 lines of context).
For ManySStuBs4J, \texttt{MORE\_SPECIFIC\_IF} and \texttt{LESS\_SPECIFIC\_IF} are among the top-5 most responsive bug patterns. For TSSM-3B these patterns do see improvement, but to a much lesser degree and, interestingly, in reverse order.
A relatively strong response can also be seen for the \enquote{more arguments} pattern, present in both datasets under slightly different names. Notably, the \enquote{opposite} pattern, that is the removal of arguments responds badly in both datasets.
Bugs that require the change of an identifier (\texttt{CHANGE\_IDENTIFIER} and \texttt{CHANGE\_IDENTIFIER\_USED}) also benefit from larger context sizes in both datasets. In general we see that bug patterns that involve or likely involve adding or changing identifiers (e.g., function names, attributes names, operands, arguments) respond strongly to context sizes. This corroborates the theory that context serves as a pool of useful ingredients.

\paragraph{Non-responding patterns}
Consistently for both datasets, operator changes range among the weakest responders, be it binary or unary operators.
We also see no or only a very weak response for boolean and numeric literal changes; in contrast to string literal changes, where there is moderate response.

\begin{figure}[tbhp]
    \centering
    \includegraphics[width=\columnwidth]{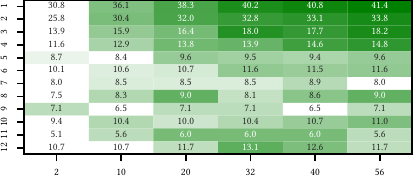}
    \caption{Heatmap visualizing the effect of context for bugs that require multiple changes to fix. As change count increases, the effect of more context diminishes. }
    \label{fig:heatmap-change}
\end{figure}

\subsubsection{Change Count}
Results show that as the number of changes increases, the effect of context steadily decreases. For six changes we still see a very weak improvement of 1.5\% from 2 to 56 lines of context; this reduces further for larger context sizes (Figure~\ref{fig:heatmap-change}).
We cannot conclusively answer why this is the case. We hypothesize that these bugs are much more difficult to fix, irrespective of context. 

\begin{answerbox}{Answer to RQ2}
The effect on repair success strongly depends on the bug type and the number of changes required for a fix. For high change counts (> 6) and certain bug types (e.g., boolean literals) increasing the context size does not substantially aid repair success.
\end{answerbox}

\subsection{Context Window Position (RQ3)}
\label{sec:rq3}
\begin{figure}[tbhp]
    \centering
    \includegraphics{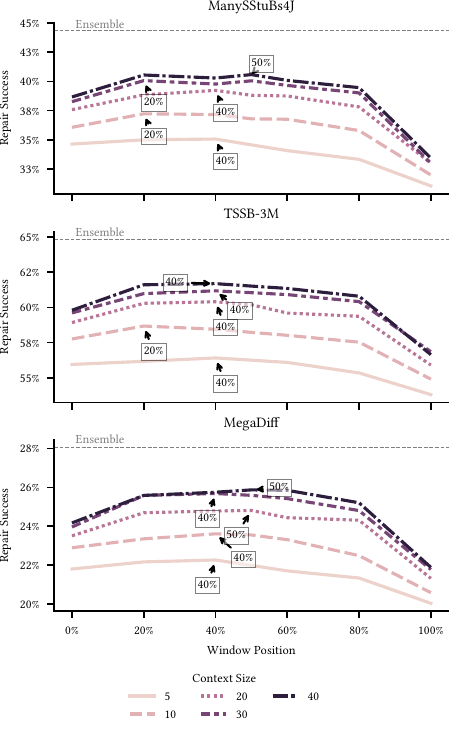}
    \caption{Repair success for different datasets, context window positions and sizes. Position of peak performance annotated. See Figure~\ref{fig:window-pos} for a visualization of window positions.}
    \label{fig:window-pos-plot}
\end{figure}

A natural choice for filling the context window is to use 50\% pre-context and 50\% post-context. We confirm this choice, although in some cases, a window slightly offset from the center might perform a bit better.
Figure~\ref{fig:window-pos-plot} shows performance across all three datasets, five context window sizes and all context window positions.

For MegaDiff, peak performance is reached at the 40\% position (i.e., 60\% of post context) for window sizes 5, 10 and 30 and 50\% for 20 and 40.
For TSSB-3M, the 40\% window was best for all context sizes, except 20. For this size, the best position lies in the other direction, at 20\%.
40\%-50\% was the best performing position for the majority of context sizes (three out of five) in the ManySStuBs4J dataset; the other two where 20\%.
For all datasets and context sizes, we see that the 0\% position (i.e., only pre-context) consistently outperforms the 100\% position (i.e., only post-context), which performs worst in all configurations. 

\begin{figure}[tbhp]
    \centering
    \includegraphics{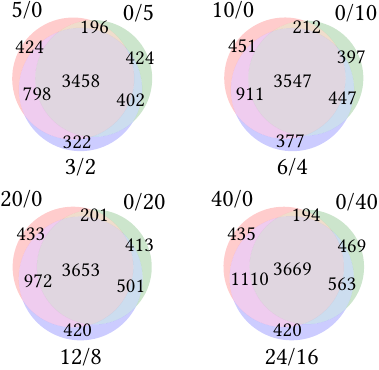}
    \caption{Are fixed bugs unique to a specific pre and post-context configuration? For four different context sizes (5, 10, 20, 40) the Venn diagrams show how many fixed bugs are unique to a specific  positional configuration (non-overlapping regions) and how many bugs could be fixed at multiple positions (overlapping regions). At the center, the number of bugs correctly fixed at all positions. Circles represent positions 100\%/0\% (\tikzcircle[fill=vennred]{3pt}), 0\%/100\% (\tikzcircle[fill=venngreen]{3pt}) and 60\%/40\% (\tikzcircle[fill=vennblue]{3pt}); bugs taken from MegaDiff.}
    \label{fig:venn}
\end{figure}

\paragraph{Complementarity}
The question naturally arises, whether, at different window positions, the model is able to fix different bugs, that is, whether different window positions are complementary to each other. To answer this question, for four of the window sizes (5, 10, 20, 40) and three window positions, we analyze how many bugs are unique (fixed only at specific window position) and how many of them are common to multiple positions. For each window size we select the extreme window positions (0\% and 100\% position) and a position close to the middle (upper median position). The results of this analysis are visualized as Venn diagrams in Figure~\ref{fig:venn}. Only 53\%-57\% of bugs are correctly fixed at all three window positions; 6\%-7\% of them are fixed by only one of the three positions. 

\paragraph{Ensembling}
Given different window positions are highly complementary, the next question that poses itself is whether models trained on window positions can be, for better performance, combined into an ensemble. When taking the five highest ranking unique (that is filtering out duplicates) predictions from the six models trained on different window positions (but the same window size) we observe a significant boost in performance. We indicate ensemble performance of the largest context window (40 lines of code) with a horizontal line in Figure~\ref{fig:window-pos-plot}. Overall, we see absolute improvements between 2\% and 3.8\% across all context sizes.

\begin{answerbox}{Answer to RQ3}
Our results suggest that peak repair success is reached at a window position of 20\%-50\%. While the best value varies across datasets and window sizes, using 60\% for pre-context and 40\% for post-context seems to be a good middle-ground. Further, different window positions allow the model to fix different bugs. This complementarity can be exploited by combining models trained on different window positions into an ensemble. 
\end{answerbox}

\subsection{Sample and Model Size (RQ4)}
\label{sec:rq4}

\begin{figure}[tbhp]
    \centering
    \includegraphics[width=\columnwidth]{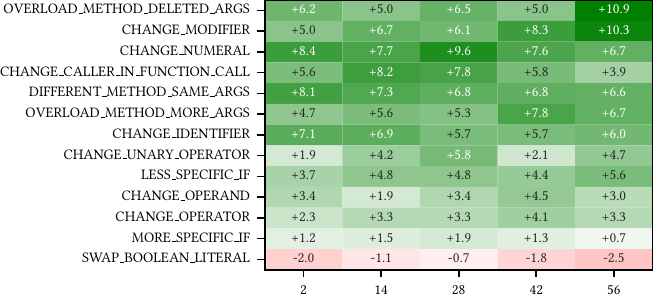}
    \caption{Difference in repair success (absolute) between the smaller 60M and the larger 220M parameter CodeT5 model for different context sizes (x) and labels (y) in ManySStuBs4J. The larger model fares better for all labels except \texttt{SWAP\_BOOLEAN\_LITERAL}.}
    \label{fig:base-comparison}
\end{figure}

\begin{figure}[tbhp]
    \centering
    \includegraphics{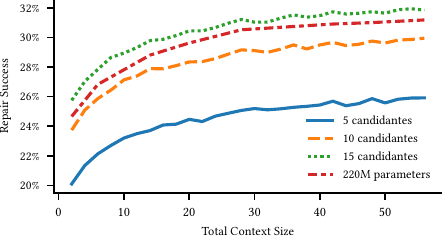}
    \caption{Repair success for different number of fix candidates (with the number of beams equal to the number of fix candidates) and the larger 220M parameter model (with 5 candidates).}
    \label{fig:samples}
\end{figure}

With increasing context, the number of possible ingredients grows (Figure~\ref{fig:rq1-overlap}). We theorize that a larger model \emph{may} be able to better exploit this wealth of ingredients. Similarly, raising the number of candidate fixes may allow the model to use more of the ingredients. If so, we would expect a \emph{disproportional} increase in performance (i.e., a widening gap) for larger context sizes. However, our results do not support this hypothesis for the tested models. 

\paragraph{Fix candidate count} When varying the number of candidates, we see that the margin between the two models remains largely constant (Figure~\ref{fig:samples}, MegaDiff).
The performance gap between 5 and 15 candidates has a mean of 5.97\% and a standard deviation of 0.16\%. Between 10 and 15 candidates, this difference decreases to 2\% (SD 0.09\%). 

\paragraph{Model Size}
Similarly, for model size we see a mostly constant gap across varying context size. 
For MegaDiff this gap has a mean and standard deviation of 4.7\% and 0.26\%, for ManySStuBs4J 5.5\% and 0.15\%. Figure~\ref{fig:base-comparison} shows a per-type comparison of performance for different bug patterns in ManySStuBs4J. We cannot discern a clear upwards trend for any pattern.
Of note, for a single bug pattern (\texttt{SWAP\_BOOLEAN\_LITERAL}) the smaller model fared better. 

\begin{answerbox}{Answer to RQ4}
As expected, both, a larger model as well as an increased number of fix candidates improves overall performance. However, we do not find convincing evidence that the gap of improvement widens at large context sizes, which would indicate a better exploitation of larger contexts. 
\end{answerbox}

\section{Implications}
\label{sec:implications}

\subsection{Opportunities of Context}

\paragraph{Context has a clear impact on performance.}
Through systematically varying the context size fed to NPR models from 2 to 56 lines, we observe \emph{relative} improvements in repair success ranging from 16\% (TSSB-3M) to 29\% (MegaDiff, ManySStuBs4J). Extending the context size from 10 to 56 lines of code still yields \emph{relative} increases between 7\% and 11.7\%.
To put these changes in perspective, this performance variation comes close to the improvements offered by some components of APR approaches, as identified through ablations. For instance, RewardRepair's semantic training improves performance by 7\%~\citet{ye2022neural}, CoCoNuT's context-awareness feature by 16\%, and the use of diagnostics in SelfAPR 37\% (all relative). The improvement is also comparable to improvements one can leverage by increasing the size of the model (\textasciitilde28\%, relative), or the number of generated samples (23\%, relative), measured on MegaDiff with the largest context size of 56 and when moving from 5 to 15 fix candidates. Importantly, the improvement obtained by increasing context appears to be orthogonal to the improvements obtained by increasing model size or number of samples.

\begin{figure}[tbhp]
    \centering
    \includegraphics{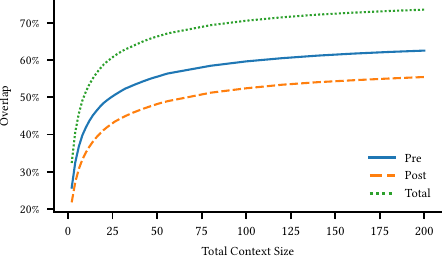}
    \caption{Overlap between identifiers in the context and the ground-truth fix for different context sizes (MegaDiff).}
    \label{fig:rq1-overlap}
\end{figure}

\paragraph{Context as a source of ingredients.} As mentioned in Section~\ref{sec:background}, context is an important source for fix ingredients~\citep{yangWhereWereRepair2021}. Furthermore, in RQ2, we identified that the changes for which the model improves most are the ones for which it needs to leverage identifiers in its context. To estimate the extent to which context ingredients could be responsible for the increased repair success we compute the overlap of identifiers in the context and the ground-truth fix. We define overlap as \( \bigl(IDs(fixed) \cap IDs(context)\bigr) / |IDs(fixed)|\),
where $IDs(c)$ denotes the set of all identifiers in $c$. Figure~\ref{fig:rq1-overlap} shows identifier overlap as a function of context size for MegaDiff. This supports the hypothesis observed from the change types in RQ2: an increased pool of ingredients, particularly identifiers, is a major factor for improved performance with larger context. Further, since both pre and post context are sources of distinct ingredients, this supports the observation in RQ3 that different context window fix different bugs. Similarly, \citet{xiaAutomatedProgramRepair2023} find that feeding post-context in addition to pre-context increases repair performance and lowers syntactic and semantic errors in the generated code.

\paragraph{Context can scale further.} Figure~\ref{fig:rq1-overlap} indicates that context can scale further: the proportion of ground-truth identifiers that enter the context continues to grow well beyond the context sizes that we studied (28 lines pre/post). At 28 lines of pre-post context, 67\% of ground truth identifiers are in context. While there are diminishing returns, this grows to 74\% if the context is extended to 200 lines of context. Even simple approaches can show appreciable benefits to leverage this additional context. By combining multiple 40-line context windows in an ensemble of models, we were able to increase repair success rate by 2 to 3.8\% (\textasciitilde5-8\% relative improvement) compared to the best 40-line window (Figure \ref{fig:window-pos-plot}). In essence, our ensemble leverages a context of up to 80 lines of context, and does so in a very straightforward fashion; other ensembling or selection strategies may further improve performance. 

\paragraph{Local context can be combined.} Finally, we note that even relatively large context sizes may not fill up the entire context window of even a modestly sized model. As mentioned in Section~\ref{sec:training}, 56 lines of context have a median token count of 592 and an upper quartile of 699. A T5 model fine-tuned with a 1024 token window comfortably fits this, leaving space for additional context enriching techniques such as SequenceR~\citep{chenSequenceRSequencetoSequenceLearning2021}, or other techniques described in Section~\ref{sec:beyond}; this is even more the case for larger models.

\subsection{Challenges with Context}

\paragraph{Context is not a panacea.} While context is helpful \emph{overall}, it is not always so. For every 10 bugs that can be reliably solved by increasing context size, there is one bug for which more context causes a regression (see Figure~\ref{fig:less-context} for an example). Moreover, an even larger proportion of bugs are sensitive to the context in ways that are hard to predict: the model may succeed to solve them at some context sizes, and not others. This is an example of the lack of robustness of neural models of code. Other work has found comparable rates of differences for both NPR~\citep{geRobustNPREvaluatingRobustnessa}, and code generation~\citep{mastropaoloRobustnessCodeGeneration2023}. This clearly calls for increased research on more robust NPR models.

We also note a silver lining: while the best performance we obtain on MegaDiff was of 26.3\% with a single model, up to 35.2\% of the bugs in MegaDiff can be fixed at one context size or another (RQ3). Combined with our finding that a straightforward ensembling technique for window sizes yielded benefits, one avenue for future research is to investigate whether ensembling models of different context sizes may help performance, and possibly alleviate the robustness issues we observed.

\paragraph{The limits of local context.} While we see further potential in extending the window size as mentioned above, there are two important caveats. First, according to Figure \ref{fig:rq1-overlap}, even when extending the context to large sizes (200 lines or more), roughly one quarter or identifiers in the ground-truth fix are not found in the context. This finding echoes other studies that found that a large proportion of method calls are non-local~\citep{karmakarJEMMAExtensibleJava2023}. Approaches that leverage other kinds of context (Section~\ref{sec:beyond}) can complement the local context. Second, while there is room to improve with the local context, whether the current crop of models can exploit it is another question: as mentioned earlier, we observed diminishing returns with more than 50 lines of context. One way forward might be models that can better generalize to longer contexts, using e.g., RoPE~\citep{suRoFormerEnhancedTransformer2022} or ALiBi~\citep{pressTrainShortTest2022}; we are unaware of such models being used in APR, beyond CodeT5's relative position embeddings.

\paragraph{Some changes are challenging, regardless of context.} Finally, while we observe some improvements when adding context, some categories of changes benefit far more than others (RQ2). In particular, the models struggled with larger changes, regardless of context size. Context size does offer some improvements for larger changes (e.g 5 or more lines), but they are slight. Similarly, while some categories of changes benefit from an increased context, for other categories context made little difference or was detrimental.

\subsection{Implications on Research Practices}

\paragraph{Context should be systematically documented.} Given the impact that context has on the performance of NPR models, clearly and thoroughly documenting the context size that was used in any experiment is crucial. However approaches from the literature are not always clear. For instance, in RewardRepair, ``the context code is considered as 10 lines of code surrounding the buggy code'' \cite{ye2022neural}. This is ambiguous: it can be interpreted either as 5 lines on each side (totalling 10), or 10 lines on each side (totalling 20).
Other work simply adds context until the input window is filled up (e.g., AlphaRepair~\citep{xiaLessTrainingMore2022a}), or uses the surrounding method as a context boundary (e.g., CoCoNuT~\citep{lutellierCoCoNuTCombiningContextaware2020}). In both cases the context size is highly variable as it depends on the size of the bug and the size of the surrounding method.
Finally, since the context window position matters, clearly specifying how the context is distributed in pre and post context is important as well.  We thus call on the community to document their choices in terms of context as clearly as possible.

\paragraph{Datasets should include enough context.} Last but not least, it is data that makes machine learning approaches possible. As mentioned in Section \ref{sec:datasets}, several NPR datasets did not forecast the need for a larger context. This includes method-level datasets such as BFP \cite{tufanoEmpiricalInvestigationLearning2018} and CoCoNuT \cite{lutellierCoCoNuTCombiningContextaware2020}, and change-level datasets with a truncated context such as ManySStuBs4J \cite{karampatsisHowOftenSingleStatement2020} and TSSM-3B\cite{richterTSSB3MMiningSingle2022}. Only MegaDiff had enough information for our study \cite{monperrusMegadiffDataset600k2021}. We hypothesize that the dearth of adequate datasets has limited studies of models with larger contexts. Indeed, we consider that re-mining the ManySStuBs4J and TSSM-3B datasets to include this additional context is one of our most important contributions. We call on the community to provide this information in future datasets, or, at a minimum, to include the information necessary for such a re-mining to take place (e.g., repository identifiers, commit hashes). 

\section{Threats to Validity}
\label{sec:limitations}
\paragraph{Bugs} Despite due diligence we cannot fully preclude software bugs in our training and evaluation scripts. 

\paragraph{Single architecture}
The scope of this work is limited to a single Transformer model (CodeT5). We chose CodeT5 as its relative position embeddings allow us to fine-tune to a larger input window size, which was a strict requirement for this study. While this model has seen widespread use in NPR, other Transformer types or neural architectures have been used (e.g., CURE~\citep{jiangCURECodeAwareNeural2021a} uses a GPT-style Transformer, CoCoNuT~\citep{lutellierCoCoNuTCombiningContextaware2020} uses convolutional networks). To which degree our results apply to them is subject to future work. With over 100 different models trained in this work, adding other architectures would have not only gone far beyond the scope of this paper but also exceeded our hardware capabilities. Instead, we focused on depth, by studying different aspects of context such as window positions, bug types and change count as well as multiple programming languages. The experiments in RQ4 show that the effects described are likely not dependent on model size or the number of candidates generated. This suggests (pending confirmation by further work) that results might extrapolate well to other Transformer types and possibly even other network types.

\paragraph{Evaluation metric and datasets}
We use exact match as an evaluation metric. Exact matching might underestimate the performance of a model, as a semantically correct fix can be expressed in multiple syntactic forms~\citep{prennerRunBugRunExecutableDataset2023}. Benchmarks such as Defects4J~\citep{justDefects4JDatabaseExisting2014} or BugsInPy~\citep{widyasariBugsInPyDatabaseExisting2020} come with test cases, which would alleviate this issue. However these benchmarks range are much smaller (hundreds of bugs). The large weight assigned to any single bug leads to concerns about noise in measurements, and associated issues with overfitting to smaller datasets, which is a concern in APR \citep{liu2021critical, noda2020experience}. We rely on two orders of magnitude more data to limit this.

\section{Conclusion}
\label{sec:conclusions}
In this work we studied the effect of local context in NPR from multiple perspectives. On multiple datasets, we first find that varying context size yields sizeable improvements (a 16--29\% relative improvement), that continue well beyond typical context local contexts used in NPR (e.g., a relative 7--11\% improvement going from 10 to 56 lines of context). Moreover, different context configurations yield different results, with straightforward ensembling techniques giving a further 5--8\% relative improvement. Our analysis of bug patterns shows that bug types involving identifiers particularly benefit from increased context. Improvements from model size and number of samples appear orthogonal to those gained from context.

Our results have multiple implications. First, there is further room for improvement as local context can be leveraged further. On the other hand, context come with challenges, particularly in terms of robustness to sometimes small variations in context. Finally, given that performance variations due to context are comparable to some improvements in the literature, we call on the community to clearly document the context they use, as well as ensuring that datasets come with enough context. We provide training and evaluation scripts as well as datasets allowing to expand this work with larger models and other neural architectures. They can be downloaded from \url{https://github.com/giganticode/out_of_context_paper_data}.

\bibliographystyle{ACM-Reference-Format}
\bibliography{main}

\end{document}